\documentstyle[aps,psfig]{revtex}
\begin{document}


\title{
Finite-Band-width Effects on the Transition Temperature and 
NMR Relaxation Rate of Impure Superconductors }
\author{Han-Yong Choi}
\address{Department of Physics, Sung Kyun Kwan University,  
Suwon 440-746, Korea}
\date{Oct. 10, 1995}
\maketitle
\begin{abstract}
We study the thermodynamic properties of impure superconductors by explicitly 
taking into consideration the finiteness of electronic bandwidths
within the phonon-mediated Eliashberg formalism.
For a finite electronic bandwidth, 
the superconducting transition temperature,
$T_c$, is suppressed by nonmagnetic impurity scatterings. 
This is a consequence of a reduction in the effective 
electron-phonon coupling, $\lambda_{eff}$.
The reduced $\lambda_{eff}$ is reflected in the observation that
the coherence peak in $1/(T_1 T)$, where $T_1$ is the
nuclear spin-lattice relaxation time and $T$ is the temperature, 
is enhanced by impurity scatterings for a finite bandwidth.
Calculations are presented for $T_c$
and $1/(T_1 T)$ as bandwidths and impurity scattering rates are varied.
Implications for doped C$_{60}$ superconductors are discussed in 
connection with $T_c$ and $1/T_1$ measurements.
 
\end{abstract}


\section{Introduction}
The Eliashberg equation  is usually solved in the limit of an infinite 
electronic bandwidth, $W$, and Fermi energy, $\epsilon_F$ 
\cite{schrieffer,allen1,pickett}. The $\epsilon_F \rightarrow \infty$ limit 
may be justified in the conventional 
superconductors where the Fermi energy
is much larger than the characteristic phonon 
frequency, $\omega_0$. The superconducting pairing occurs
mainly within a region of width $\omega_0$ around the Fermi surface,
and for $\epsilon_F \gg \omega_0$, it makes no difference whether we take
$\epsilon_F$ finite or infinite.
There are, however, superconducting materials where $\epsilon_F$ is 
comparable with $\omega_0$:
For fullerene superconductors, $\omega_0 \approx
0.05 \-- 0.2$ eV, $\epsilon_F \approx 0.05 \-- 0.2$ eV, and
$\omega_0 /\epsilon_F \sim 1$ \cite{hebard}. 
It is, therefore, of great interest to study
how the superconducting properties are modified as the bandwidth and
Fermi energy are reduced. 
This is an important as well as difficult problem which concomitantly calls for
a reinvestigation of the Migdal theorem and Coulomb pseudopotential $\mu^*$ 
\cite{schrieffer,allen1,bcs}.
The Migdal theorem ensures that the phonon vertex corrections are smaller
than those terms included in the theory of superconductivity 
by the factor of $\omega_0 /\epsilon_F $,
hence may be neglected for conventional low temperature
superconductors with $\epsilon_F \gg \omega_0$. 
The Coulomb repulsion, $\mu$, is reduced to $\mu^* = \mu/[1+
\mu \log(\epsilon_F /\omega_0 )] \approx 0.1 \-- 0.2$,
owing to the retardation effect.
If $\omega_0 /\epsilon_F \sim 1$, however,
then the phonon vertex correction should
be important, and the Coulomb repulsion will not be
reduced, $\mu^* \approx \mu \sim 1 $, which will almost always kill
superconductivity. 

We will not, however, try to investigate these complications in this
paper. We will instead assume that superconductivity results 
within the framework 
of Eliashberg theory, and will investigate how the strong-coupling
Eliashberg theory is modified due to a finite bandwidth 
and what the consequences are of the modification.
This problem was first treated by Zheng and 
Bennemann \cite{zheng} in the context of fullerene superconductors, 
who calculated the pressure dependence of the transition temperature,
$T_c$, for doped 
fullerenes within the strong-coupling Eliashberg formalism. 
Their calculations with the finite bandwidth explicitly included
agree well with the experimental observations.
In the different context of a $non$-$phonon$ superconductivity, 
Marsiglio \cite{marsi1} investigated the dependence of $T_c$ on
nonmagnetic and magnetic impurities including the effects of a finite 
bandwidth. He found that the nonmagnetic impurities reduce $T_c$ 
for finite $W$.
This is interesting in view of the recent debate over the Abrikosov 
and Gorkov (AG) theory of impure superconductors \cite{agd,maki1}. 
Kim and Overhauser \cite{kim1} 
pointed out that the AG theory, if evaluated literally within 
the $non$-$retarded$ weak-coupling Bardeen-Cooper-Schrieffer (BCS) 
framework \cite{bcs}, implies a substantial reduction 
of $T_c$ by nonmagnetic impurities in an apparent 
contradiction to the Anderson's theorem \cite{anderson}
and to experimental observations on conventional
low temperature superconductors \cite{allen1}. 
It is now understood that the AG theory is in accord with the Anderson's
theorem if treated within the Eliashberg theory taking fully into account
the $retarded$ nature of a pairing interaction,
whereas the criticism by Kim and Overhauser should be valid for 
a non-retarded pairing interaction \cite{radtke,fay}.
The problem of retardedness is directly related with a cutoff in the
frequency or momentum summation. By working with a more realistic 
finite bandwidth Eliashberg theory,
with cutoffs both in the frequency and momentum summations, 
we will be able to understand the recent controversy on the AG theory 
more clearly as will be discussed later.

In the AG theory of impure 
superconductors of infinite bandwidths, the thermodynamic properties
are independent of the impurity scattering rates
because there exists a scaling relation 
between pure and impure superconductivity \cite{agd,maki1}.
This simple relation, however, breaks down if finite bandwidths
are taken into consideration as will be discussed below.
Therefore, the thermodynamic properties of dirty superconductors are no 
longer independent of impurity scattering rates when the finite 
bandwidths are explicitly taken into account. 
Two of the thermodynamic properties, 
transition temperature, $T_c$, and nuclear spin-lattice 
relaxation rate, $T_1^{-1}$, are studied in this paper
within the phonon-mediated Eliashberg theory. 
When an electronic bandwidth is finite, $T_c$ is suppressed 
as impurity scattering rate, $\tau^{-1}$, is increased. 
The rate of $T_c$ suppression by impurities is determined by the electronic 
bandwidth and strength of electron-phonon coupling, and in the limit of
$\epsilon_F \rightarrow \infty$, $T_c$ is independent of impurity scattering 
rate in agreement with the Anderson's theorem.
The impurity suppression of $T_c$ is $not$ a consequence of a time reversal 
symmetry breaking, but follows dynamically from the modified Eliashberg 
equation of a finite electronic bandwidth, as is reflected in the fact that 
nuclear spin-lattice relaxation rate $T_1^{-1}$ is 
rather enhanced by the impurity scatterings.
As the electronic bandwidth is reduced, the available electronic states
to and from which quasi-particles can be scattered are restricted,
and the effective electron-phonon coupling constant, $\lambda_{eff}$, 
consequently, is decreased. The reduced $\lambda_{eff}$ implies
decreased $T_c$ and enhanced NMR coherence peak in $(T_1 T)^{-1}$.

There are several factors that affect the NMR
coherence peak of ${(T_1 T)}^{-1}$ \cite{nmrrev,choi1}. 
The suppression of NMR coherence peak may be attributed to 
(a) gap anisotropy/non $s$-wave pairing 
of superconducting phase \cite{nmrrev},
(b) strong-coupling phonon damping \cite{nakamura,allen2}, 
(c) paramagnetic impurities in samples \cite{maki1,nmrrev}, and/or
(d) strong Coulomb interaction \cite{koyama} 
such as paramagnon/antiparamagnon effects \cite {hasegawa}. 
The nonmagnetic impurity scatterings, on the other hand, have no influence 
on ${(T_1 T)}^{-1}$ for conventional superconductors of infinite 
bandwidth \cite{agd,maki1,anderson}, other than the smearing of
gap anisotropy.
This is because (a) there exists a simple 
scaling relation between the gap and renormalization functions of pure
and impure superconductors of infinite bandwidth, 
and (b) in the expression for ${(T_1 T)}^{-1}$, 
the numerator and denominator are of the same powers in 
renormalization function, as will be detailed below.
We found that $T_c$ is reduced and the NMR coherence peak is enhanced 
as $\tau^{-1}$ is increased for a finite bandwidth,
because the effective $\lambda_{eff}$ is reduced.
This interpretation is consistent with the previous study of paramagnetic
transition metals by MacDonald \cite{macdon}.
In considering the effects of a finite bandwidth
of the $d$ electrons on the mass enhancement,
MacDonald found the reduced bandwidths 
cause a reduction in the mass enhancement
due to electron-paramagnon interaction.
The reduced $m_{eff}$ means reduced $\lambda_{eff}$
in agreement with the present work. 

This paper is organized as follows:
In Sec. II, we will present the Eliashberg equation 
on imaginary frequencies for impure superconductors with finite bandwidths.
Within the formalism, we will also discuss 
the Anderson's theorem including the scaling relation between the pure 
and impure superconductivity.
We can solve the Eliashberg equation either in the imaginary frequency
to obtain the gap function, $\widetilde{\Delta} (ip_n )$, and renormalization 
function, $\widetilde{Z} (ip_n )$,
or in the real frequency to obtain $\widetilde{\Delta} (\omega )$ 
and $\widetilde{Z} (\omega )$. 
It is, however, much easier to solve the Eliashberg equation 
in the imaginary frequency. We therefore carried out the
$T_c$ calculations in the imaginary frequencies. Detailed numerical
calculations of as well as qualitative discussions on $T_c$ will be 
presented in detail in Sec. III. A brief comment on the
theory of $T_c$ of impure superconductivity will be made in view of
recent debates on the topic.
For calculating $1/T_1$, we need 
$\widetilde{\Delta} (\omega )$ and $\widetilde{Z} (\omega )$ 
in real frequencies. It is more efficient to solve the Eliashberg 
equation in the imaginary frequency and perform analytic continuations
to real frequency than to solve it in real frequency.
Using the iterative method for analytic continuations \cite{msc} 
extended to finite bandwidths, 
we calculate the nuclear spin-lattice relaxation rates 
$1/T_1$ as the bandwidths, electron-phonon
couplings and impurity scattering rates are varied.
The results of these calculations will be presented in Sec. IV.
Finally, we will
summarize our results and give some concluding remarks in Sec. V.

\section{Eliashberg Theory of Finite Bandwidth Superconductors}
The electron-phonon interaction is local in space and retarded in time.
Consequently, it's momentum dependence is weak and neglected in the 
isotropic Eliashberg equation, but the frequency dependence is 
important and fully included.
The isotropic Eliashberg equation 
in the imaginary frequency including the finite electronic bandwidth 
and impurity scatterings on an equal footing is written as 
\cite{allen1,pickett}
\begin{eqnarray}
\Sigma (ip_n ) = -\frac{1}{\beta} \sum_m \lambda(n-m) g(ip_m)
- \frac{1}{2\pi \tau} g(ip_n),
\label{self}
\end{eqnarray}
where
\begin{eqnarray}
\lambda(n-m) &=& N_F \int_{0}^{\infty} d\Omega~ \frac {2\Omega ~
\alpha^2 F(\Omega)}{\Omega^2 +(p_n -p_m )^2 },
\nonumber \\
g(ip_m) = -\sum_{\vec k'} \tau_3 G( ip_m , \vec{k'}) \tau_3 &=&
\int_{-W/2}^{W/2} d\epsilon_k \frac 
{i\widetilde{W}_m + (\epsilon_k -\mu +\chi_n ) 
\tau_3 - \widetilde{\phi}_m \tau_1} 
{ \widetilde{W}_m^2 +(\epsilon_k -\mu +\chi_n )^2 
+\widetilde{\phi}_m^2 }.
\label{lambda}
\end{eqnarray}
Here, $N_F$ is the density of states at the Fermi level, 
$\tau_i$'s the Pauli matrices operating in the Nambu space, 
$i \widetilde{W}_n \equiv ip_n \widetilde{Z}_n$, 
$\widetilde{\phi}_n \equiv \widetilde{\Delta}_n \widetilde{Z}_n$, 
where $p_n$ is the Matsubara frequency given by
$p_n = \pi (2n+1) / \beta$, $\beta = 1/{k_B T}$, and $\mu$ is the chemical
potential, which should not be confused with the Coulomb repulsion.
$\widetilde{Z}_n \equiv \widetilde{Z}(ip_n) $,
$ \widetilde{\Delta}_n \equiv \widetilde{\Delta}(ip_n) $ and 
$\chi_n \equiv \chi (ip_n) $ are, respectively,
the renormalization function, gap function and energy shift, 
when analytically
continued to real frequency. A tilde on a variable denotes that it is 
renormalized by impurity scatterings.
From Eq.\ (\ref{self}), we obtain the following three coupled equations:
\begin{eqnarray}
i\widetilde{W}_n &=& ip_n + \frac{1}{\beta} \sum_m \lambda(n-m)~ 
\frac{(\widetilde{\theta}_m^1 -\widetilde{\theta}_m^2 ) i\widetilde{W}_m} 
{\sqrt{\widetilde{W}_m^2 +\widetilde{\phi}_m^2 } } +\frac{1}{2\pi\tau}
\frac{(\widetilde{\theta}_n^1 -\widetilde{\theta}_n^2 ) i\widetilde{W}_n} 
{\sqrt{\widetilde{W}_n^2 +\widetilde{\phi}_n^2 } },
\nonumber \\
\widetilde{\phi}_n &=& \frac{1}{\beta} \sum_m \lambda(n-m)~ 
\frac{(\widetilde{\theta}_m^1 -\widetilde{\theta}_m^2 ) \widetilde{\phi}_m} 
{\sqrt{\widetilde{W}_m^2 +\widetilde{\phi}_m^2 } } +\frac{1}{2\pi\tau}
\frac{(\widetilde{\theta}_n^1 -\widetilde{\theta}_n^2 ) \widetilde{\phi}_n} 
{\sqrt{\widetilde{W}_n^2 +\widetilde{\phi}_n^2 } },
\nonumber \\
\chi_n &=& \frac{1}{\beta} \sum_m \lambda (n-m) \log \biggl [ \frac
{\cos(\widetilde{\theta}_m^1 )}{\cos(\widetilde{\theta}_m^2 )} \biggr ] 
+\frac{1}{2\pi\tau}
\log \biggl [ \frac {\cos(\widetilde{\theta}_n^1 )}
{\cos(\widetilde{\theta}_n^2 )} \biggr ],
\label{elia1}
\end{eqnarray}
where 
\begin{equation}
\widetilde{\theta}_n^1 = \tan^{-1} \biggl [ \frac {W/2 -\mu +\chi_n }
{ \sqrt{ \widetilde{W}_n^2 +\widetilde{\phi}_n^2 } } \biggr ], ~
\widetilde{\theta}_n^2 = \tan^{-1} \biggl [ \frac {-W/2 -\mu +\chi_n }
{ \sqrt{ \widetilde{W}_n^2 +\widetilde{\phi}_n^2 } } \biggr ].
\label{ang}
\end{equation}
This coupled equation should be solved simultaneously with the following
constraint of number conservation which determines $\mu$:
\begin{equation}
n_f = \frac{1}{2} +\frac {1}{\beta} \sum_{n,\vec k} Tr \Bigl [ 
G(ip_n, \vec{k} ) \tau_3 \bigr ] e^{i \delta},
\end{equation}
where $n_f$ is the band filling factor, $n_f = N_e /N_a$, 
$N_e$ the number of electrons, $N_a$ the number of available states 
including the spin degeneracy
factor, and $\delta$ is a positive infinitesimal.
We took the Coulomb pseudopotential
$\mu^* = 0$ in Eq.\ (\ref{elia1}) for simplicity.

For the half-filled case of $n_f =1/2$, 
Eq.\ (\ref{elia1}) is greatly simplified 
because $\chi_n$ and $\mu$ vanish identically.
We have
\begin{eqnarray}
i\widetilde{W}_n &=& ip_n + \frac{1}{\beta} \sum_m \lambda(n-m)~ 
\frac{2\widetilde{\theta}_m i\widetilde{W}_m} 
{\sqrt{\widetilde{W}_m^2 +\widetilde{\phi}_m^2 } } 
+\frac{1}{\pi\tau} 
\frac{\widetilde{\theta}_n i\widetilde{W}_n} 
{\sqrt{\widetilde{W}_n^2 +\widetilde{\phi}_n^2 } },
\nonumber \\
\widetilde{\phi}_n &=& \frac{1}{\beta} \sum_m \lambda(n-m)~ 
\frac{2\widetilde{\theta}_m \widetilde{\phi}_m } 
{\sqrt{\widetilde{W}_m^2 +\widetilde{\phi}_m^2 } } 
+\frac{1}{\pi\tau} 
\frac{\widetilde{\theta}_n \widetilde{\phi}_n} 
{\sqrt{\widetilde{W}_n^2 +\widetilde{\phi}_n^2 } },
\label{elia2}
\end{eqnarray}
where
\begin{equation}
\widetilde{\theta}_n = \tan^{-1} \biggl [ 
\frac {W} {2 \sqrt{\widetilde{W}_n^2 +\widetilde{\phi}_n^2 } } \biggr ].
\label{ang2}
\end{equation}
This is the Eliashberg equation for the half-filled case including
finite bandwidths and impurity scatterings.
For discussions on the Anderson's theorem, it is convenient to rearrange
Eq.\ (\ref{elia2}), moving the last terms of the right hand side to 
the left, as follows:
\begin{eqnarray}
i\widetilde{W}_n \zeta_n &=& ip_n + \frac{1}{\beta} \sum_m \lambda(n-m)~
\frac{2 \widetilde{\theta}_m i\widetilde{W}_m} 
{\sqrt{\widetilde{W}_m^2 +\widetilde{\phi}_m^2 } },
\nonumber \\
\widetilde{\phi}_n \zeta_n &=& \frac{1}{\beta} \sum_m \lambda(n-m)~
\frac{2 \widetilde{\theta}_m \widetilde{\phi}_m} 
{\sqrt{\widetilde{W}_m^2 +\widetilde{\phi}_m^2 } },
\label{elia3}
\end{eqnarray}
where $\zeta_n = 1 - \frac{\widetilde{\theta}_n}{\pi \tau} 
\frac{1} {\sqrt{\widetilde{W}_n^2 
+\widetilde{\phi}_n^2 } }$.
If we multiply both the numerator and denominator of the right 
hand side of Eq.\ (\ref{elia3}) by $\zeta_m$ and
identify $\widetilde{W}_m \zeta_m = W_m$ and 
$\widetilde{\phi}_m \zeta_m = \phi_m$, 
where untilded variables simply imply that they are for pure superconductors,
we obtain 
\begin{eqnarray}
iW_n &=& ip_n + \frac{1}{\beta} \sum_m \lambda(n-m)~
\frac{2 \widetilde{\theta}_m iW_m} {\sqrt{W_m^2 +\phi_m^2 } },
\nonumber \\
\phi_n &=& \frac{1}{\beta} \sum_m \lambda(n-m)~
\frac{2 \widetilde{\theta}_m \phi_m} {\sqrt{W_m^2 +\phi_m^2 } }.
\label{eliap}
\end{eqnarray}
Note that the correct equation that describes
the pure superconductors is obtained
from Eq.\ (\ref{elia2}) by taking the limit $\tau \rightarrow \infty$, which
is just Eq.\ (\ref{eliap}) 
with $\widetilde{\theta}_m$ replaced by $\theta_m$.

When the bandwidth is infinite, we have 
$\widetilde{\theta}_m$ = $\theta_m$ = $\pi/2$,
and Eq.\ (\ref{eliap}) describes both
pure ($W_n$ and $\phi_n$) and 
impure ($\widetilde{W}_n$ and $\widetilde{\phi}_n$) superconductivity.
They are related by the following scaling relation:
\begin{eqnarray}
\widetilde{W}_n &=& \eta_n W_n,~
\widetilde{\phi}_n = \eta_n \phi_n,~
\eta_n = 1 + \frac{1}{2 \tau \sqrt{W_n^2 +\phi_n^2 } },
\nonumber \\
{\rm or,}~~ \widetilde{\Delta}(\omega) &=& \Delta(\omega),~
\widetilde{Z}(\omega) = Z(\omega) +\frac{i}{2 \tau \sqrt{\omega^2 -
\Delta(\omega)^2} }.
\label{scale}
\end{eqnarray}
This constitutes a proof of the Anderson's theorem that the transition
temperature and other transport properties remain unchanged under 
nonmagnetic impurity scatterings for infinite $W$.
When the bandwidth is finite, 
Eq.\ (\ref{eliap}) is not the correct equation describing 
pure superconductivity. Consequently, the Anderson's theorem
does not hold in this case, and we expect that the thermodynamic properties,
such as transition temperature $T_c$, 
will change as the impurity scatterings are
introduced. We will turn to this topic in the following section.

\section{Transition Temperature}
Before we solve the Eliashberg equation of Eq.\ (\ref{elia2}),
let us first try an approximate solution to understand what to expect
from detailed numerical calculations.
From Eq.\ (\ref{ang2}), $\widetilde{\theta}_n \approx \frac{\pi}{2} -\frac
{2}{W} \sqrt{\widetilde{W}_n^2 +\widetilde{\phi}_n^2}$. We plug this into
Eq.\ (\ref{eliap}), and take $T = T_c$ so that $\phi_m = 0$, to obtain
\begin{eqnarray}
Z_n &=& 1 +\frac{\pi}{p_n \beta} \sum_m \lambda(n-m) \frac{p_m}{|p_m|}
\left( 1 -\frac{4}{\pi W} \left| \widetilde{W}_m \right| \right),
\nonumber \\
1 &=& \frac{\pi}{\beta} \sum_m \lambda(n-m)
\frac{1}{|W_m|} \left( 1 -\frac{4}{\pi W} \left|
\widetilde{W}_m \right| \right).
\label{appeqn}
\end{eqnarray}
We take $\lambda(\epsilon -\epsilon') = \lambda$ for 
$ |\epsilon | \leq \omega_0$ and $ |\epsilon'| \leq \omega_0$,
and 0 otherwise, which is a standard weak-coupling approximation.
Then, taking the advantage of the scaling relation of Eq.\ (\ref{scale}),
we obtain
\begin{eqnarray}
Z_n = 1+\lambda \left( 1-\frac{1}{\pi \tau \epsilon_F} \right)
-\frac{2\lambda |p_n|}{\pi \epsilon_F} Z_n
= \frac{1 +\lambda \left(1-1/\pi\tau\epsilon_F \right)}
{1 +2\lambda |p_n|/\pi\epsilon_F}.
\label{zn}
\end{eqnarray}
The second equation of Eq.\ (\ref{appeqn}), after
carrying out the summation over Matsubara frequency $ip_m$ using
\begin{eqnarray}
\frac{1}{\beta} \sum_m F(ip_m) &=& 
\int \frac{dz}{2 \pi i} f(z) F(z),
\nonumber \\
F(ip_n) &=& \frac{1}{\pi} \int d\epsilon \frac{1}{\epsilon -ip_n}
Im \Bigl[ F(\epsilon) \Bigr],
\end{eqnarray}
can be written as
\begin{equation}
1 = \pi \int \frac{dz}{2\pi i} f(z)~
\frac{1}{\pi} \int d\epsilon \frac{1}{\epsilon -z}
Im \biggl[ \frac{i\lambda}{|\epsilon Z|} -\frac{2}{\pi \epsilon_F} 
\Bigl(1 +\frac{i}{2 \tau |\epsilon Z|} \Bigr) \biggr],
\end{equation}
where $f(z) = 1/(1+e^{\beta z})$ is the Fermi distribution function.
Then,
\begin{eqnarray}
1 = -&\lambda_{eff}& \int_{-\omega_0}^{\omega_0} d\epsilon
\frac{f(\epsilon)}{\epsilon} 
= \lambda_{eff} \log \Bigl[ 1.13 \omega_0 /T_c \Bigr],
\nonumber \\
&\lambda_{eff} &= \frac{\lambda \left( 1- 1/\pi \tau \epsilon_F \right)}
{1+\lambda \left( 1- 1/\pi \tau \epsilon_F \right) }.
\label{leff}
\end{eqnarray}
From this, we find
\begin{eqnarray}
T_c = 1.13 \omega_0 ~e^{-1/\lambda_{eff}} \approx
T_{cp} -\frac{1}{\pi \tau \epsilon_F \lambda} T_{cp},
\label{apptc} 
\end{eqnarray}
where $T_{cp} = 1.13 \omega_0~ e^{-(1+\lambda)/\lambda} $
is the transition temperature for pure superconductor of
a given bandwidth. We note that $T_{cp} $ is equal to 
the transition temperature of infinite bandwidth, $T_{c0}$, in the
approximate treatment that replaces $Z(\epsilon)$ by $ Z(0)$. More detailed
numerical calculations, however, yield that $T_{cp} \leq T_{c0}$
as shown in Fig. 1.
Eq.\ (\ref{apptc}) clearly shows that impurities are pair-breaking when 
the bandwidth is
finite in agreement with Marsiglio \cite{marsi1}.
$T_c$ is reduced not because the time reversal symmetry is broken
as is the case with magnetic impurities, but 
because the effective electron-phonon coupling constant is
decreased by impurity scatterings for finite bandwidth
as can be seen from Eq.\ (\ref{leff}).
For $\epsilon_F \rightarrow \infty$, Eq.\ (\ref{apptc}) implies that
$T_c$ is not changed as was discussed in the previous section.

It seems appropriate to comment here on the recent debate on the AG theory 
of impure superconductivity. This, as already pointed out by 
Radtke \cite{radtke}, 
stems from using non-retarded interaction of weak-coupling
scheme \cite{kim1}.
For this, let us rewrite Eq.\ (\ref{eliap}) with $\widetilde{\theta}_m
=\pi/2$ for infinite bandwidth,
before the integral over $\epsilon_k$.
\begin{eqnarray}
\phi_n = \frac{\pi}{\beta} \sum_m \int_{-\infty}^{\infty} d\epsilon_k ~
\lambda(n-m) \frac{\widetilde{\phi}_m}
{\widetilde{W}_m^2 +\widetilde{\phi}_m^2 +\epsilon_k^2}.
\end{eqnarray}
Then, we have, for $T= T_c$
\begin{equation}
1 = \frac{\pi}{\beta} \sum_m \int_{-\infty}^{\infty} d\epsilon_k ~
\lambda(n-m) \frac{\eta_m}{\eta_m^2 p_m^2 +\epsilon_k^2}.
\end{equation}
If we proceed the same way leading to Eq.\ (\ref{apptc}), we have
$T_c = T_{c0}$ for $\epsilon_F \rightarrow \infty$.
Kim and Overhauser \cite{kim1}, however, exchanged the range of summation 
between $\epsilon_k$ and $ip_n$, as did AG, 
and evaluated the resulting equation exactly.
This leads to a contradiction to the Anderson's theorem
and experimental observations, as was pointed out by them.
The origin of this contradiction is clear. 
This procedure amounts to neglecting the retarded nature of
electron-phonon interaction, because $\lambda (n-m)$ now becomes
frequency independent, which implies
an instantaneous interaction in time.

The impurity suppression of $T_c$ we discuss in this paper
should be distinguished from the previous studies \cite{marsi1,kim1}.
Kim and Overhauser found that the nonmagnetic impurities suppress $T_c$ for 
a non-retarded pairing interaction. 
We point out that $T_c$ is suppressed by nonmagnetic impurities
for a retarded pairing interaction also if the bandwidth is finite.
On the other hand, 
Marsiglio found that nonmagnetic impurities suppress $T_c$ of
finite bandwidth superconductors for non-phonon pairing interactions.
It is now understood that his results are due to both
the finiteness of $W$ and non-phonon nature of pairing interactoins.
For a pairing interaction not of the form of Eq.\ (\ref{elia2}), the proof of
the Anderson's theorem of Sec. II does not go through. Therefore, 
nonmagnetic impurity scatterings can suppress the transition temperature.

\begin{figure}
\centerline{\psfig{file=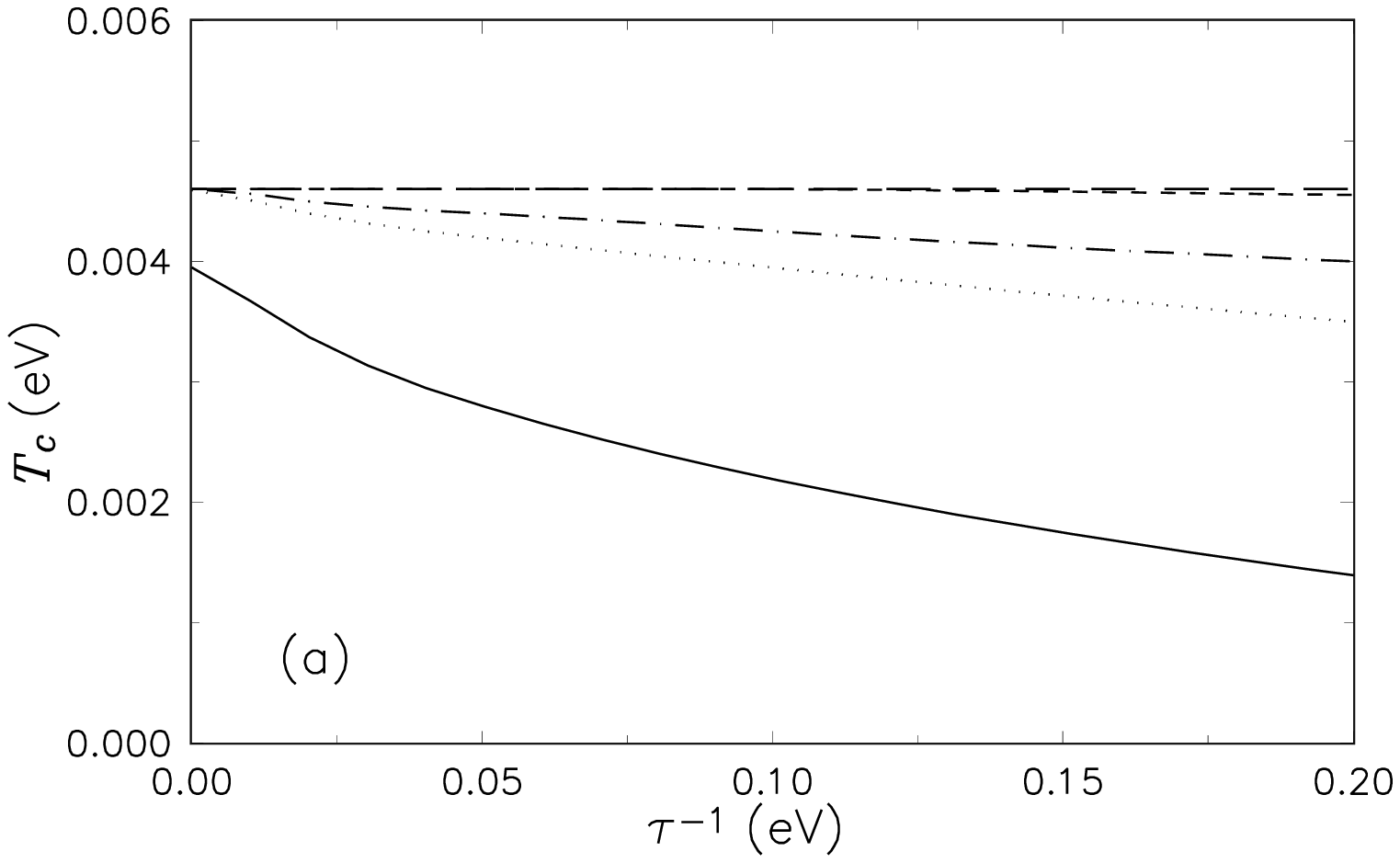,width=0.5\linewidth}}
\centerline{\psfig{file=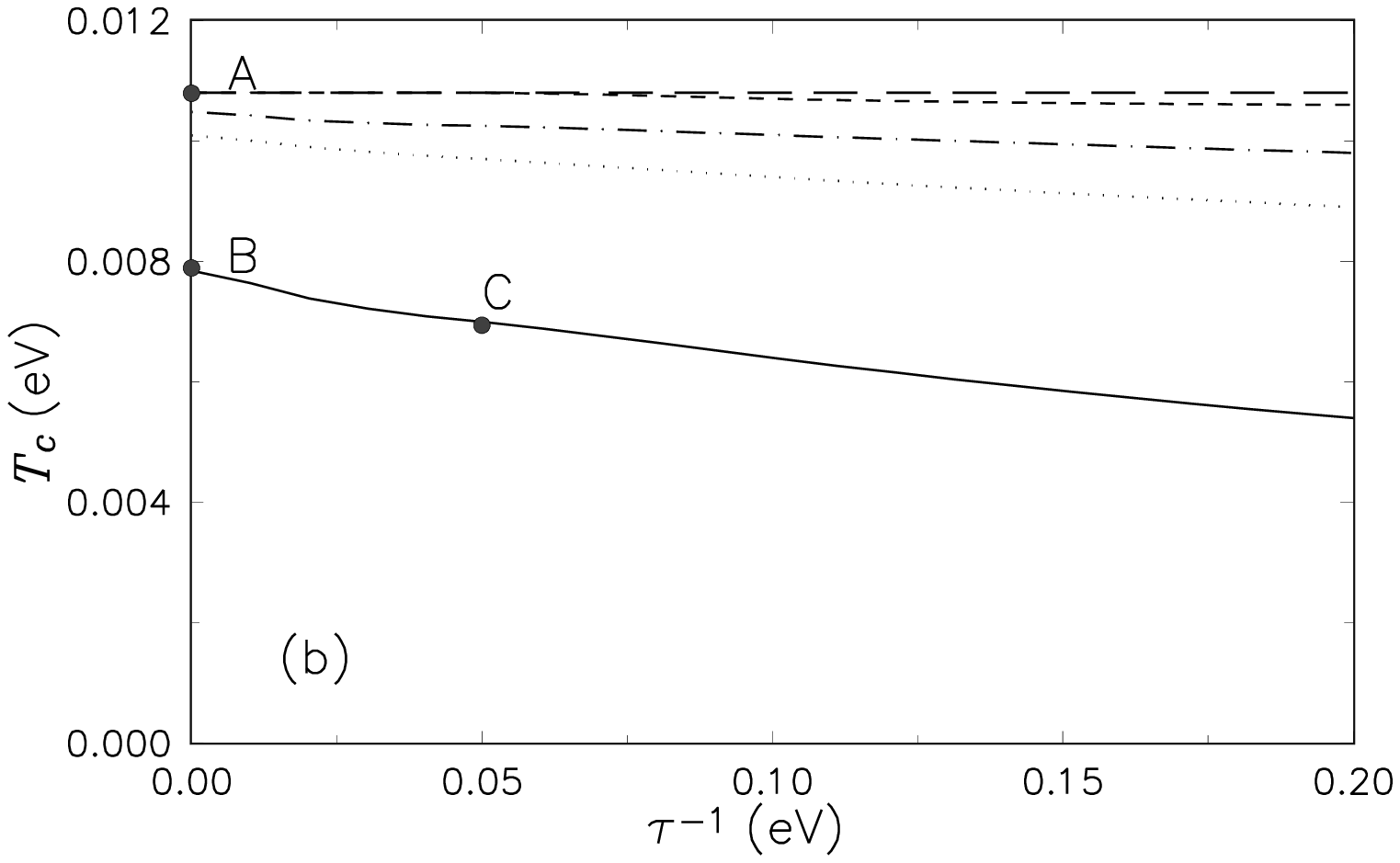,width=0.5\linewidth}}
\caption{
Transition temperature, $T_c$, as a function of impurity scattering rate,
$\tau^{-1}$, for half-filled finite bandwidth superconductors,
as calculated from Eq. 6.
Fig. 1(a) is for $N_F \alpha^2$ = 0.02 eV and Fig. 1(b) for 0.05 eV.
The solid, dotted, dot-dashed, short-dashed, and long-dashed curves
represent, respectively, bandwidth $W$ = 0.1, 0.5, 1, 5, and 10 eV.
The dots labeled as A, B, and C in Fig. 1(b) are selected for
calculating NMR relaxation rates shown in Fig. 3.
$T_c$ is decreased by impurity scatterings for finite bandwidth
supercondcutors, while it is unchanged for those with infinite bandwidths.
The rate of $T_c$ suppression by impurities is larger for
narrower bandwidth superconductors.
}
\label{fig1}
\end{figure}

The qualitative discussion above is well verified in the detailed
numerical calculations.
In Fig. 1, we show the transition temperature $T_c$, 
calculated from Eq.\ (\ref{elia2}), as a function of
impurity scattering rate $\tau^{-1}$ for several bandwidths.
We took following Bickers {\it et al.} \cite{bickers}
\begin{eqnarray}
F(\Omega) = \left \{
\begin{array}{ll}
\frac{1}{R} \biggl[ \frac{1}{ (\Omega-\omega_0)^2 +\Gamma^2 } 
-\frac{1}{\Gamma_c^2 +\Gamma^2} \biggr] &
\text{for}~ |\Omega-\omega_0| \leq \Gamma_c , \\
0 & \text{otherwise}, 
\end{array}
\right.
\label{fomega}
\end{eqnarray}
with $\omega_0$ = 0.05 eV, $\Gamma_c = 3\Gamma$ = 0.015 eV.
$R$ is a normalization constant to make
$\int_0^{\infty} d\Omega~F(\Omega) = 1$.
Using 200 Matsubara frequencies, self-consistency is reached
within a few tens of iterations at a given temperature, except for 
temperatures close to $T_c$.
The solid, dotted, dot-dashed, short-dashed, and long-dashed curves
correspond, respectively, to $W$ = 0.1, 0.5, 1, 5, and 10 eV.
We considered half-filled cases for simplicity, so that the Fermi energy
$\epsilon_F = W/2$.
Fig. 1(a) and (b) are for $N_F \alpha^2$ = 0.02 and 0.05 eV, respectively.
The long-dashed curves representing $W$ = 10 eV, 
are indistinguishable from infinite bandwidth curves.
As we expected from the qualitative analysis above, the impurity
suppression of the transition temperature is more pronounced for
narrower bandwidths.
As the bandwidth becomes wider, however, the rate of $T_c$ suppression
by impurity scatterings is smaller until $\epsilon_F \approx 1$, beyond
which we are almost in the infinite bandwidth limit where $T_c$ is independent
of the impurity scattering rate in accordance with the Anderson's theorem.

We wish to consider how other transport properties of finite bandwidth
superconductors, NMR relaxation rate for example, are altered
in the following section.
If the reduction in $T_c$ is due to time reversal symmetry breaking,
the NMR coherence peak below $T_c$ will be reduced.
If, on the other hand, it is due to reduction of 
the effective electron-phonon
coupling as is the case we consider here, we expect that the NMR
coherence peak will be enhanced. This is because
the strong-coupling effects cause phonon dampings and
reduce the coherence peak as was discussed.

\section{Nuclear Spin-Lattice Relaxation Rate}
To calculate nuclear spin-lattice relaxation rate $T_1^{-1}$,
we need to perform analytic continuation 
to obtain the gap and renormalization functions on real
frequency, $\widetilde{\Delta}(\omega)$ and $\widetilde{Z}(\omega)$, 
from those on imaginary frequency,
$\widetilde{\Delta}(ip_n)$ and $\widetilde{Z}(ip_n)$.
It was carried out via the iterative method by Marsiglio, Schossmann, 
and Carbotte (MSC) \cite{msc}.
The MSC equation which relates the gap and renormalization functions 
on imaginary frequency with those on real frequency, extended
to half-filled finite bandwidth impure superconductors, is given by
\begin{eqnarray}
\widetilde{Z}(\omega) &=& 
1 +\frac{i}{\omega} \int_{-\infty}^{\infty} d\Omega ~
\alpha^2 F(\Omega) \frac{2 \widetilde{\theta}(\omega-\Omega)~ 
(\omega -\Omega)} {\sqrt{(\omega-\Omega)^2 -\widetilde{\Delta}
(\omega-\Omega)^2} } \Bigl[ N(\Omega) +f(\Omega-\omega) \Bigr] 
\nonumber \\
&+& \frac{1}{\beta \omega} \sum_{n\geq 0} \frac{2 \widetilde{\theta}_n ip_n} 
{\sqrt{p_n^2 +\widetilde{\Delta}_n^2} }
\Bigl[ \lambda(\omega -ip_n) -\lambda(\omega +ip_n) \Bigr]
+ \frac{i}{\pi \tau \omega} 
\frac{\widetilde{\theta}(\omega) \omega}
{\sqrt{\omega^2 -\widetilde{\Delta}(\omega)^2} },
\nonumber \\
\widetilde{\Delta}(\omega) &=& 
\frac{i}{\widetilde{Z}(\omega)} \int_{-\infty}^{\infty} d\Omega ~
\alpha^2 F(\Omega) \frac{2 \widetilde{\theta}(\omega-\Omega)~ 
\widetilde{\Delta}(\omega -\Omega)}
{\sqrt{(\omega-\Omega)^2 -\widetilde{\Delta}(\omega-\Omega)^2} } 
\Bigl[ N(\Omega) +f(\Omega-\omega) \Bigr] 
\nonumber \\
&+& \frac{1}{\beta \widetilde{Z}(\omega)} \sum_{n\geq 0} 
\frac{2 \widetilde{\theta}_n \widetilde{\Delta}_n} 
{\sqrt{p_n^2 +\widetilde{\Delta}_n^2} }
\Bigl[ \lambda(\omega -ip_n) +\lambda(\omega +ip_n) \Bigr]
+ \frac{i}{\pi \tau \widetilde{Z}(\omega) } 
\frac{\widetilde{\theta}(\omega) \widetilde{\Delta}(\omega)}
{\sqrt{\omega^2 -\widetilde{\Delta}(\omega)^2} },
\label{msc}
\end{eqnarray}
where
\begin{eqnarray}
\lambda(\omega \pm ip_n) = N_F \int_{0}^{\infty} d\Omega~ \frac {2\Omega ~
\alpha^2 F(\Omega)}{\Omega^2 -(\omega \pm ip_n )^2 }.
\end{eqnarray}
This equation is solved iteratively for 
$\widetilde{Z}(\omega)$ and $\widetilde{\Delta}(\omega)$,
taking $\widetilde{Z}_n$ and
$\widetilde{\Delta}_n$, solution to the Eliashberg equation on
Matsubara frequencies of Eq.\ (\ref{elia2}), as an input,
until self-consistency is reached.
We show in Fig. 2 the gap function $\widetilde{\Delta} (\omega)$ as a function of
$\omega$ at $T$ = 0.001 eV. 
We took the impurity scattering rate $\tau^{-1}$ = 0, and
the phonon spectral function $\alpha^2 F (\Omega)$
as given by Eq.\ (\ref{fomega}) with $N_F \alpha^2 = 0.05$ eV
which corresponds to Fig. 1(b). 
Fig. 2(a) is for an infinite bandwidth, 
and (b) is for bandwidth $W$ = 1 eV. The solid and dashed lines,
respectively, stand for real and imaginary parts of the gap function.
The obtained results for $W \rightarrow \infty$, where previous studies
are available, are in good agreement with the published 
data \cite{bickers}.

\begin{figure}
\centerline{\psfig{file=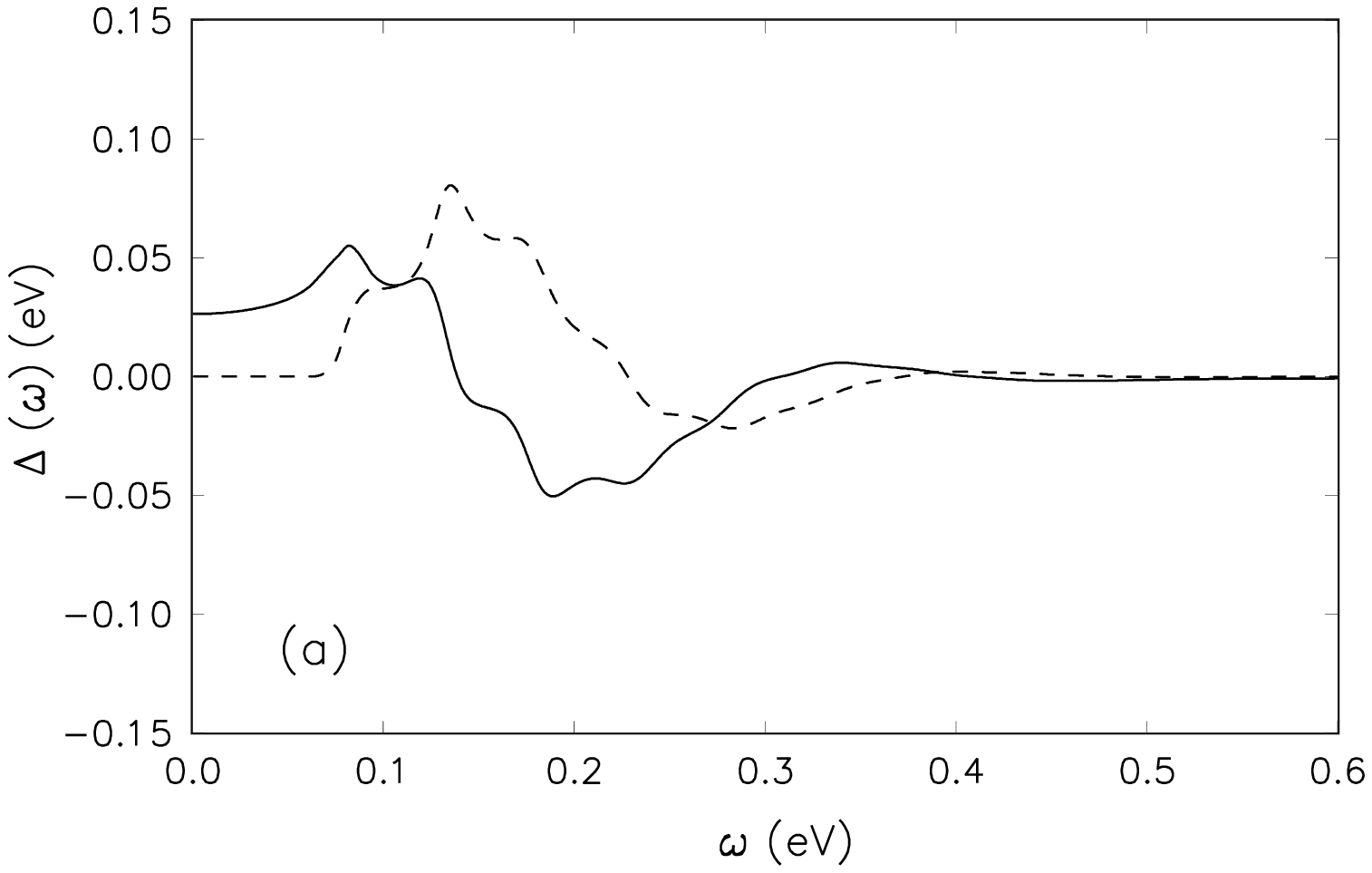,width=0.5\linewidth}}
\centerline{\psfig{file=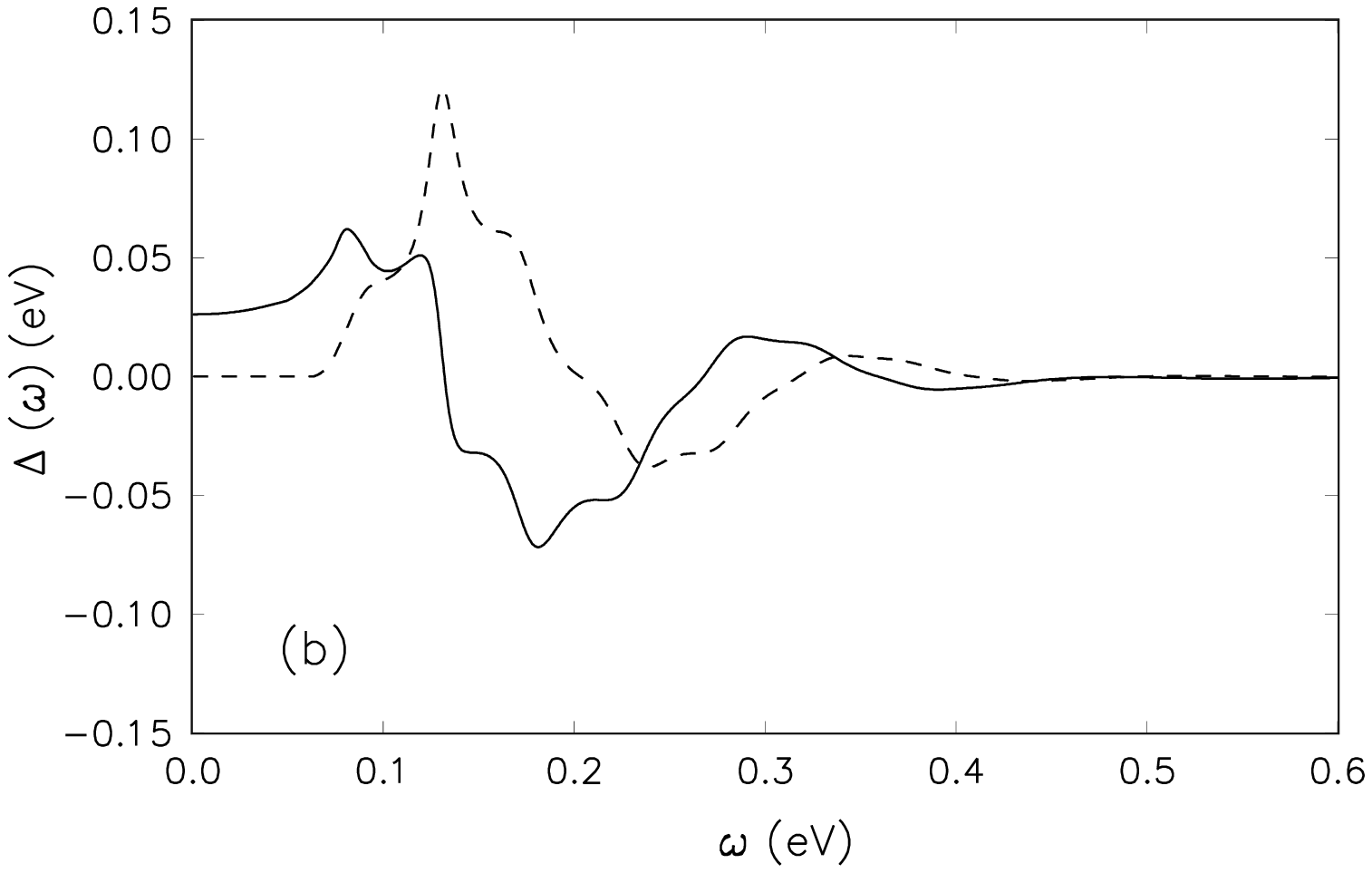,width=0.5\linewidth}}
\caption{
The gap function, $\Delta (\omega)$, as a function of $\omega$ at
$T$ = 0.001 eV. We took $\tau^{-1}$ = 0 and
$N_F \alpha^2$ = 0.05 eV, which corresponds to Fig. 1(b).
The solid
and dashed lines, respectively, represent the real and imaginary parts
of the gap.
Fig. 2(a) is for an infinite bandwidth, and (b) is for $W$ = 1 eV.
}
\label{fig2}
\end{figure}

The nuclear spin-lattice relaxation rate 
$T_1 ^{-1}$ is given by \cite{moriya}
\begin{eqnarray}
\frac{1}{T_1} = \lim_{\omega \rightarrow 0} 
\frac{1}{1 - e^{-\beta \omega} } \sum_{\vec{q}} \Big| A_q \Big|^2
Im \Bigl[ \chi_{+-} (\omega + i \delta ,\vec{q}) \Bigr],
\label{t1}
\end{eqnarray}
where $A_q$ is a form factor related with the conduction electron 
wavefunctions, and $\chi_{+-} ( \omega , \vec{q}) $ is
a spin-spin correlation function at frequency $\omega$ and momentum transfer
$\vec{q}$.
The impurity scatterings are included in the self-energy of renormalized
Green's function.
For a finite bandwidth, it is easy to derive
\begin{eqnarray}
\frac{1}{T_1 T} &\propto& \int_0^{\infty} d\epsilon 
\frac{\partial f(\epsilon) }{\partial \epsilon}~
\biggl\{ \biggl[ Re \biggl(
\frac{\epsilon~ \widetilde{\theta}(\epsilon)}
{\sqrt{\epsilon^2 -\widetilde{\Delta}(\epsilon)^2}} \biggr) \biggr]^2
+ \biggl[ Re \biggl(
\frac{\widetilde{\Delta}(\epsilon)~ \widetilde{\theta}(\epsilon)}
{\sqrt{\epsilon^2 -\widetilde{\Delta}(\epsilon)^2}} \biggr) \biggr]^2 
\biggr\}.
\label{t1t-s} \\
&\rightarrow & \int_{\Delta}^{\infty} d\epsilon 
\frac{\partial f}{\partial \epsilon}~ \biggl\{
\frac{\epsilon^2 +\Delta^2}{\epsilon^2 -\Delta^2} \biggr\},
\label{t1t-w}
\end{eqnarray}
where the $\widetilde{\Delta}(\epsilon)$ and 
$\widetilde{\theta}(\epsilon)$ are obtained by solving
Eqs.\ (\ref{elia2}) and (\ref{msc}) iteratively.
The standard strong-coupling expression for $T_1^{-1}$ given, for example,
by Fibich \cite{fibich} can be obtained by putting
$\widetilde{\theta} = \pi/2$ for infinite bandwidth.
Eq.\ (\ref{t1t-w}) follows for infinite bandwidth weak-coupling limit.

Before we present the detailed numerical calculations, let us first
analyze the expression for $(T_1 T)^{-1}$ of 
Eq.\ (\ref{t1t-s}) qualitatively.
For $T > T_c$, $\widetilde{\Delta}$ = 0 in Eq.\ (\ref{t1t-s}), and
\begin{eqnarray}
\widetilde{\theta}(\epsilon) = \tan^{-1} \biggl[
\frac{i W}{2 \epsilon \widetilde{Z}(\epsilon)} \biggr] =
\left \{
\begin{array}{ll}
i \tanh^{-1} \biggl[ 
\frac{W}{2\epsilon \widetilde{Z}(\epsilon)} \biggr] &
{\rm for}~ \epsilon \geq \frac{W}{2 \widetilde{Z}}, \\
\frac{\pi}{2} +i \tanh^{-1} \biggl[ 
\frac{2\epsilon \widetilde{Z}(\epsilon)}{W} \biggr] &
{\rm for}~ \epsilon \leq \frac{W}{2 \widetilde{Z}}.
\end{array}
\right.
\label{ang3}
\end{eqnarray}
Taking $\widetilde{Z}(\epsilon) \approx 1$ for simplicity, we have
$Re \Bigl[ \widetilde{\theta}(\epsilon) \Bigr] = 
\pi/2$ for $\epsilon \leq W/2$,
and 0 for $\epsilon > W/2$. Then,
\begin{eqnarray}
\frac{1}{T_1 T} \propto \int_0^{W/2} &d\epsilon&
\frac{\partial f(\epsilon)}{\partial \epsilon} =
-\frac{1}{2} \tanh \Bigl[ \textstyle{\frac{1}{2} }\beta W \Bigr],
\nonumber \\
{\rm or,}~~ \frac{(T_1 T)_n}{(T_1 T)_s} &=& \frac {\tanh \Bigl[ W/2T \Bigr]}
{\tanh \Bigl[ W/2T_c \Bigr]}.
\end{eqnarray}
$(T_1 T)_n /(T_1 T)_s $, therefore, decreases as the temperature is
increased, which is contrasted with the constant value of 1 for the
infinite bandwidth case.
The decrease of $(T_1 T)_n /(T_1 T)_s $ as the temperature is
increased above $T_c$ was observed in the $^{51}$V NMR study of
V$_3$Si superconductors by Kishimoto {\it et al.} \cite{kishmoto}. 
This observation was interpreted in terms of 
narrow bandwidths in accord with the present work.
The range of integration from 0 to $W/2$ is just  
what we may expect intuitively.
The $T < T_c$ region, on the other hand, is difficult to analyze without
detailed information on $\widetilde{\Delta}(\epsilon)$, because the
height of NMR coherence peak is mainly determined by the magnitude
of imaginary part of $\widetilde{\Delta}(\epsilon)$ in the Eliashberg
formalism. We may expect, however, that the coherence peak will be
enhanced as $W$ is decreased and $\tau^{-1}$ is increased, because
$\lambda_{eff}$ is reduced.

\begin{figure}
\centerline{\psfig{file=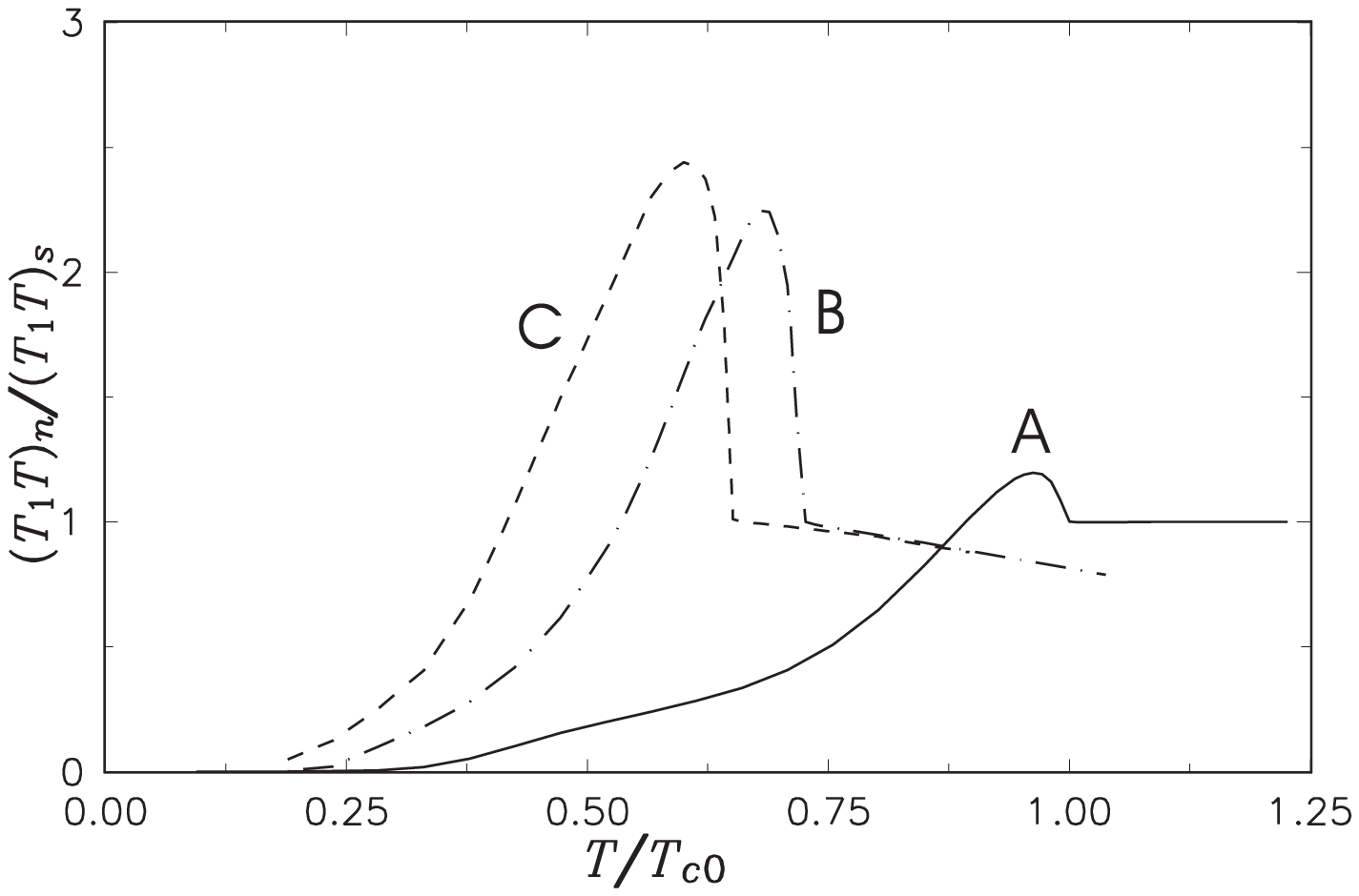,width=0.5\linewidth}}
\caption{
The normalized nuclear spin-lattice relaxation rate by the normal state
Korringa value, $(T_1 T)_n /(T_1 T)_s$, as a function of the reduced
temperature, $T/T_{c0}$, where $T_{c0}$ is the critical temperature
of infinite bandwidth case, with $N_F \alpha^2$ = 0.05 eV.
The solid, dot-dashed, and dashed curves, labeled, respectively,
as A, B, and C, are computed for $W$ and $\tau^{-1}$ equal to
10 and 0, 0.1 and 0, and 0.1 and 0.05 eV, as can be read off
from their counterparts in Fig. 1(b).
The normalized relaxation rates show progressively
enhanced peaks as one goes from A to B to C.
This can be understood in terms of the effective electron-phonon coupling
constant alone, because large $\lambda_{eff}$ suppresses
NMR coherence peak due to strong-coupling phonon dampings.
As $W$ is reduced and $\tau^{-1}$ is increased,
$\lambda_{eff}$ is reduced. The computed $\lambda_{eff}$ for A, B, and C
are 1.67, 0.63, and 0.60, respectively.
}
\label{fig3}
\end{figure}

In Fig. 3, we show the normalized NMR relaxation rate by the normal
state Korringa value, $(T_1 T)_n /(T_1 T)_s $, calculated
from Eqs.\ (\ref{elia2}), (\ref{msc}) and (\ref{t1t-s})
as a function of $T/T_{c0}$, where $T_{c0}$ is the critical temperature
for infinite bandwidth case.
We took the phonon spectral function $\alpha^2 F(\Omega)$ as given by 
Eq.\ (\ref{fomega}) with $N_F \alpha^2 = 0.05$ eV, which corresponds
to Fig. 1(b). 
We selected 3 sets of $W$ and $\tau^{-1}$ values, 
labeled as A, B, and C in Fig. 1(b), for $T_1^{-1}$
calculations. 
$W$ and $\tau^{-1}$ of A, B, and C are, respectively, in unit of eV, 
10 and 0, 0.1 and 0, and 0.1 and 0.05, as can be read from Fig. 1(b).
Note that as one goes from A to B to C, the normalized NMR relaxation rates
have progressively enhanced peaks, and $T_c$ is reduced accordingly.
These results are straightforward to interpret, as already
explained before.
The computed values of $\lambda_{eff}$ 
for A, B, and C are 1.67, 0.63, and 0.60,
respectively. As $\lambda_{eff}$ is decreased, $T_c$ should be reduced and
NMR coherence peak should be enhanced, because there is
no time reversal symmetry breaking in the present problem.
The solid curve of A, having $\lambda_{eff}$ = 1.67, has substantially
reduced coherence peak, in agreement with the previous works
\cite {nakamura,allen2}.
Note also that the normalized NMR relaxation rates for finite bandwidths
decrease as $T$ is increased beyond $T_c$ as expected.

\section{Summary and Concluding Remarks}
In this paper, we have investigated the effects of a finite bandwidth 
on the thermodynamic properties of impure superconductors
within the framework of phonon-mediated Eliashberg theory. We found that
the transition temperature and NMR coherence peak are suppressed 
and enhanced, respectively, by impurity scatterings
when the finiteness of bandwidths is explicitly taken into consideration.
These results can be understood in terms of 
reduced effective electron-phonon coupling $\lambda_{eff}$.
The motivation for this work was, in part, the observation that 
the phonon frequency and the Fermi energy are comparable and
a substantial disorder is present in the fullerene superconductors
\cite{hebard}.
The NMR coherence peak in $(T_1 T)^{-1}$ was found absent for
doped fullerenes \cite{tycko,sasaki}. 
We wish to point out that the present theory is $not$ concerned with 
why the NMR coherence peak is absent for a given material. 
The present theory shows that
if the disorder is increased for a finite bandwidth superconductor,
its transition temperature should be reduced and coherence peak should 
be enhanced, respectively, compared with those of a clean material.
In view of our results, it will be very interesting to systematically
investigate how the transition temperature and NMR coherence peak behave 
as the degree of disorder is varied for doped C$_{60}$.

In A$_3$C$_{60}$, almost all other experiments than NMR 
relaxation rates seem to point to a phonon-mediated $s$-wave pairing 
\cite{hebard}. Also, due to the orientational disorder \cite{stephens}, 
the Fermi surface anisotropy is not strong enough
to suppress the coherence peak \cite{mele}. 
The present study shows that
a quite strong electron-phonon coupling of $\lambda_{eff} \approx 2$ is 
needed to suppress the NMR coherence peak in agreement with
Nakamura $et~al.$ \cite{nakamura}, and Allen and Rainer \cite{allen2}.
The $\lambda_{eff} \approx 2$ seems too large for doped
fullerenes since the far infrared reflectivity measurements 
of DeGiorge {\it et al.} \cite{degiorge}
show that $2\Delta/k_B T_c \approx 3.44 - 3.45$, 
a classic weak-coupling value.
Because there are no magnetic impurities in A$_3$C$_{60}$, 
the absence of NMR coherence peak is still to be understood.
Stenger {\it et al.} \cite{stenger} 
suggested that the applied magnetic field is responsible for
the suppressed NMR coherence peak. Their explanation, however, seems
to be more a puzzle than an answer. 
According to the Eliashberg theory,
the energy of applied magnetic field should be at least $\hbar \omega
\approx 0.1 \Delta$ to suppress the coherence peak. 
The magnetic field in their $^{13}$C NMR experiment corresponds to
$\hbar \omega \approx 10^{-5} \Delta$.
Such a small energy scale in the coherence peak suppression
is really a puzzle.
We are currently investigating the strong Coulomb interaction effects
with a paramagnon approximation \cite{hasegawa}. 
We suspect that the strong Coulomb interaction may be responsible 
for suppressed NMR coherence peak in doped fullerenes.

This work was supported by Korea Science and Engineering Foundation 
(KOSEF) through Grant No. 951-0209-035-2,
and by the Ministry of Education through Grant No. BSRI-95-2428.

\end{document}